\documentclass[journal]{IEEEtran}
\usepackage{graphicx}
\usepackage{epstopdf}
\usepackage[cmex10]{amsmath}
\usepackage{amssymb}
\usepackage{amsthm}
\DeclareMathOperator{\Tr}{Tr}
\DeclareMathOperator{\Var}{Var}
\DeclareMathOperator{\Cov}{Cov}
\DeclareMathOperator{\VEC}{vec}
\DeclareMathOperator{\blockdiag}{blockdiag}

\usepackage{bm,cite}
\usepackage{algorithm}
\usepackage{algpseudocode}
\usepackage{tikz}
\usetikzlibrary{arrows}
\usepackage{verbatim}
\usepackage{pgfplots}
\pgfplotsset{compat=1.13}
\usetikzlibrary{positioning,calc}
\usetikzlibrary{plotmarks}
\newlength\figureheight
\newlength\figurewidth

\begin{document}

\title{Cooperative Channel Estimation for Coordinated Transmission with Limited Backhaul}

\author{Qianrui~Li,~\IEEEmembership{Member,~IEEE,}
        David~Gesbert,~\IEEEmembership{Fellow,~IEEE}
        and~Nicolas~Gresset,~\IEEEmembership{Member,~IEEE}% <-this % stops a space

\thanks{Part of this work has been presented at the International Conference on Communication (ICCW 2015) Small Cell and 5G Networks  (SmallNets) Workshop, London, June 2015. }
\thanks{Q. Li and N. Gresset are with Mitsubishi Electric R\&D Centre Europe, 35708 Rennes, France. (e-mail:{q.li,n.gresset}@fr.merce.mee.com) Q. Li was also with Communication System Department, EURECOM during this work.}
\thanks{D. Gesbert is with Communication System Department, EURECOM, 06410 Biot,France. (e-mail:gesbert@eurecom.fr)}
}

\maketitle

\begin{abstract}
Obtaining accurate global channel state information (CSI) at multiple transmitter devices is critical to the performance of many coordinated transmission schemes. Practical CSI local feedback often leads to noisy and partial CSI estimates at each transmitter. With rate-limited bi-directional backhaul, transmitters have the opportunity to exchange few CSI-related bits to establish global channel state information at transmitter (CSIT). This work investigates possible strategies towards this goal. We propose a novel decentralized algorithm that produces minimum mean square error (MMSE)-optimal global channel estimates at each device from combining local feedback and information exchanged through backhauls. The method adapts to arbitrary initial information topologies and feedback noise statistics and can do that with a combination of closed-form and convex approaches. Simulations for coordinated multi-point (CoMP) transmission systems with two or three transmitters exhibit the advantage of the proposed algorithm over conventional CSI exchange mechanisms when the coordination backhauls are limited.

\end{abstract}


% \fi
\IEEEpeerreviewmaketitle

\section{Introduction}

The acquisition of global CSI at multiple devices that engage in the coordinated transmission \cite{marsch2011coordinated} is crucial to the system performance in many cooperative transmission techniques such as CoMP (or network multiple-input and multiple-output, abbreviated to network MIMO) \cite{5594708}, coordinated beamforming and scheduling \cite{5463229,6129540}, and interference alignment \cite{cadambe2008interference}.
In currently envisioned mobile system evolutions, two trends, namely centralized and decentralized co-exist. In the centralized
architecture, transmitter cooperation is supported by  the so-called Cloud Radio AccessNetwork (C-RAN) where baseband processing is pushed into core networks  with very high rate optical backhaul links \cite{checko2015cloud}.  In contrast, when optical-enabled C-RAN is too expensive or not suitable for the network deployments (e.g., heterogeneous backhaul featuring wireless links, mobile relays, flying relays, flexible on-demand low-cost deployment, temporary hot-spot coverage), a higher degree of decentralization is required. Hence, local information based processing is desired so as to keep the backhaul overhead low and channel measurement's time relevance high. Other deployment paradigms calling for decentralized transmitter cooperation with explicit limited information exchange constraints include so-called Dynamic Radio Access Network (Dynamic RAN) \cite{metisD6} and inter-operator spectrum sharing.

Transmitter cooperation with limited backhaul has been studied for a long time. In \cite{4601071}, they have analyzed the multi-cell processing performance with finite-capacity backhaul using information theory tools. The work in \cite{4698016,Bhagavatula2011Adaptive} also consider the sum rate maximization of the CoMP system with constrained backhaul. A quantization scheme for CSI sharing under finite-capacity backhaul assumption implementing interference alignment is proposed by \cite{Rezaee2013CSIT}. However, such designs employ CSI quantizations, with no regard for the {\em statistical properties} of the local information already existing at the transmitter receiving the information and ignoring the potential benefits of {\em correlated initial channel estimates} available at the transmitters.

In this work, we recast the problem into a more general and systematic decentralized channel estimation problem with side information \cite{7247161}. In this setup, each transmitter (TX) starts by acquiring an initial CSI estimate from any local feedback mechanism. Interestingly, such mechanisms can be of arbitrary nature, encompassing scenarios such as the current long-term evolution (LTE) release, where each base station can only acquire CSI related to a subset of the users which are served by that base station. Other scenarios can be also accounted such as broadcast feedback (feedback is overheard by all TXs within a certain distance) and hierarchical feedback designs, where in the latter, some of the TXs (e.g. so-called "master base stations") are endowed by design with a greater amount of CSI compared with surrounding "slave" TXs \cite{6853644}. More generally, the initial CSI structure may exhibit an arbitrary level of accuracy as well as spatial correlation (from TX to TX). A general and not previously addressed problem can then be formulated as follows: Given the arbitrary initial CSI structure and the finite information exchange capability between the TXs, what are the reasonable strategies for cooperation (among TXs) for the purpose of generating CSI with high-enough quality at each TX? An interesting side question is how much information should flow in each direction for every TX pair when backhaul links are subject to a global bidirectional rate constraint.

In this paper, the information exchange through capacity-limited backhaul is modeled via a fixed rate quantization scheme. The final CSI estimate is generated based on an MMSE criterion, and involves a suitable combining of the initial local CSI feedback and the backhaul-exchanged information acquired from other TXs. A difficulty in this problem lies in the fact that the optimization of the information exchange schemes and that of the  CSI combining scheme are fundamentally coupled. Nevertheless our contribution reveals that the two optimization steps can be undertaken jointly.

Clearly, the problem of cooperative channel estimation is rooted in the information theoretic framework of network vector quantization \cite{fleming2004network} and lossy source coding with side information  (i.e. Wyner-Ziv coding \cite{wyner1976rate}). When more than two TXs are involved in the cooperation, it becomes related to the problem of multiple-source compression with side information at the decoder \cite{del2008distributed}. The information theoretical bound \cite{yamamoto1980source} and asymptotically bound-achieving quantizer design for Wyner-Ziv coding \cite{zamir2002nested,liu2006slepian,rebollo2004wyner} are well analyzed. However, some of those designs only valid under specific cooperation topology assumption and the complexity is high for real implementation. In this paper we are interested in reasonable complexity, practically implementable optimization algorithms for which Wyner Ziv coding schemes can serve as useful benchmarks.

In this work, we propose a novel optimization framework, referred as {\em coordination shaping}, which addresses the above problems under a wide range of noise and initial CSI feedback design. Our specific contributions include:

\begin{itemize}
 \item A joint optimized quantization and information combining scheme allowing to produce MMSE-optimal global CSI at all nodes of the network.  The quantizer minimizes a weighted distortion measure where the weight (quantization shaping) matrix is optimized as a function of the distributions of CSI quality across the cooperating TXs. The final CSI estimate at each TX  linearly combines the initial and exchanged CSI. A key finding is that the quantization shaping matrices and linear combining weights can be optimized jointly by a convex program.
 \item For the case of two TXs, our proposed algorithm can even outperforms the Wyner-Ziv transform
 coding algorithm \cite{rebollo2004wyner} in the low rate region, while the performance asymptotically achieves the Wyner-Ziv bound in
 the high resolution regime.
 \item The proposed algorithm works for multi-transmitter cooperation scenarios as well, hence offering a generalized low-complexity implementation of Wyner-Ziv coding based schemes.
 \item The proposed framework is exploited to find the optimal coordination bit allocation in the case of global bidirectional rate constraint.
\end{itemize}

\section{System model and problem description}
\label{sec:sysmod}
We consider a communication system with $K$ TXs and $L$ receivers (RX). The cooperative TXs could be base stations attempting to serve  receiving terminals in a cooperative fashion. There exists many cooperative transmission strategies, generally requiring the availability of some global CSI at each TX \cite{5594708,5463229,6129540,cadambe2008interference}. Although the actual choice of the transmission scheme (joint MIMO precoding, interference alignment, coordinated scheduling, coordinated resource allocation, etc.) may affect the CSI reconstruction problem at the TX side, such a question is left for further work while this paper focuses instead on the general problem of producing the best possible global CSI at each and every TX in a non discriminatory manner. The impact of our channel estimation framework on the overall system performance is however evaluated in Section \ref{sec:nuperf} for a particular example of network-MIMO enabled system.

Let's assume that each TX~$i, \forall i=1,\dots,K$ is equipped with $M$ transmit antennas while each RX~$j, \forall j=1,\dots,L$ is equipped with $N$ receive antennas. The propagation channel between TX~$i$ and RX~$j$ is denoted as $\mathbf{H}_{ji}\in \mathbb{C}^{N\times M}$. The full network-wide MIMO channel is $\mathbf{H} \in \mathbb{C}^{NL\times MK}$ with:
\begin{equation*}
\begin{array}{lcl}
 \mathbf{H}&=&\left[\begin{array}{lcl}
 \mathbf{H}_{11}&\ldots&\mathbf{H}_{1K}\\
 \vdots&\ldots&\vdots\\
 \mathbf{H}_{L1}&\ldots&\mathbf{H}_{LK}
 \end{array}\right].
 \end{array}
\end{equation*}
We consider frequency-flat Rayleigh fading channels. $\mathbf{h}=vec(\mathbf{H})\sim\mathcal{CN}(\mathbf{0},\mathbf{Q}_{\mathbf{h}})$, where the vector $\mathbf{h}$ is the vectorized version of full network-wide MIMO channel $\mathbf{H}$ and $\mathbf{Q}_{\mathbf{h}}$ is an arbitrary multi-user channel covariance matrix.

\subsection{Distributed CSI model}
\label{ssec:DCSImod}

For the CSI model, we assume that each TX acquires an initial estimate of the global channel state from an arbitrary pilot-based, digital or analog feedback mechanism. Similar to the CSI model used in \cite{wagner2012}, the CSI made initially available at TX~$i$ is a noisy one. More generally, the CSI imperfection is TX-dependent, giving rise to a distributed CSI model as initially introduced in \cite{6248219}. Let
\begin{equation}\label{eq:chmod}
\hat{\mathbf{H}}^{(i)}=\mathbf{H}+\mathbf{E}^{(i)},
\end{equation}
where $\hat{\mathbf{H}}^{(i)}\in \mathbb{C}^{NL\times MK}$ is a CSI estimate for $\mathbf{H}$ at TX~$i$. $\mathbf{E}^{(i)}\in \mathbb{C}^{NL\times MK}$ is the estimation error seen at TX~$i$. Hence, the {\em estimates} at various TXs can be correlated through ${\bf H}$.
%The TX dependence of the CSI estimates for possible the same quantity constitutes a major deviation from classically studied centralized CSI settings.
The channel independent $\mathbf{E}^{(i)}$ satisfies $\VEC(\mathbf{E}^{(i)})=\mathbf{e}^{(i)}\sim\mathcal{CN}(\mathbf{0},\mathbf{Q}_i)$. The values of $\mathbf{Q}_i$ entries depend on the mean channel gains for each TX-RX pair as well as the CSIR feedback bit allocation.
The errors terms seen at different TXs are assumed independent, i.e, $\mathbb{E}\{\mathbf{e}^{(i)}\cdot {\mathbf{e}^{(j)}}^H\}=\mathbf{0}, \forall i\neq j$. Throughout this work, the channel statistics $\mathbf{Q}_{\mathbf{h}}$ and all error statistics $\mathbf{Q}_i, \forall i=1,\dots,K$ are assumed to be known at every TX by virtue of slow statistical variations. In practical system, those statistics can be calculated by the pilot based channel estimation procedure. Thanks to the slow variation, the signaling overhead and delay constraints for statistic information exchange is much easier to fulfill than those for instantaneous channel. Therefore, we assume that statistics are perfectly shared among TXs while sharing instantaneous CSI is constrained by limited backhaul.
%This assumption implies that the sharing of channel related statistics among TXs is accurate. It is not affected by the limited backhaul constraints while sharing the instantaneous CSI is not the case.

Note that the value of this CSI model lies in the fact that it is quite general, including diverse scenarios ranging from local to global information structures. The problem of CSI exchange between cooperating base stations is currently an important topic in $3$GPP discussions in order to evaluate the real benefits of CoMP. Although different feedback schemes such as broadcast feedback and hierarchical feedback are not supported yet in the current LTE/LTE-A standards, they are promising CSI feedback technologies for $5$G, where the system should become more UE-centric, at least in some of the envisioned $5$G scenarios. Different CSI structures and feedback schemes can be described using the aforementioned CSI model. For example, in the conventional LTE downlink channel estimation scenario, the channel estimation is performed in FDD mode by each Base Station (TX) sending pilots to the users (RXs). Each user will feedback its downlink CSI to its associated Base Station only. This gives rise to a strongly local initial CSI at each TX. In the scenario of {\em broadcast feedback}, a terminal feeds back its downlink CSI to all overhearing TXs, thus providing global CSI estimates at all TXs where it matters most. Still, in this case too, the quality of the initial CSI estimates remains TX-dependent due to small scale and large scale effects on the uplink. More general various degrees of {\em locality} for the initial CSI are completely captured in the structure of $\mathbf{Q}_i$. A channel component with corresponding noise variance equal to zero in $\mathbf{Q}_i$ will indicate perfect knowledge of that coefficient at TX~$i$. A channel component with smaller corresponding noise variance than another indicates that the local initial CSIT for this channel component is more accurate. Finally, $\mathbf{Q}_i=\mathbf{0}, \forall i$ refer to the perfect centralized CSI case, which renders backhaul-based CSI exchange superfluous.

\subsection{Limited rate coordination model}
\label{ssec:lrcmod}

Let's consider the transmitter devices are equipped with rate-limited bi-directional communication links over which they can exchange a finite amount of CSI related information. Note that we only allow a {\em single shot} of coordination which consumes $R_{ki}$ bits of communication from TX~$k$ to TX~$i$ for all cooperating pairs $(k,i)${\em simultaneously}. The problem is now to optimally exploit this coordination capability so as to acquire the best possible global channel estimate at each TX.

\begin{figure}[tp!]
\centering
\includegraphics[width=1\columnwidth]{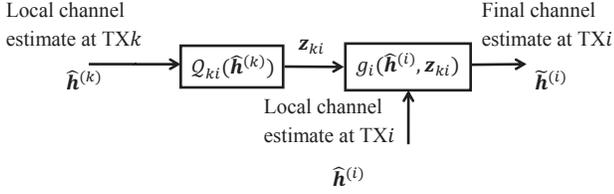}
\caption{Decentralized cooperative channel estimation across two TXs.}\label{fig:shapedcoordproc}
\end{figure}

In Fig. \ref{fig:shapedcoordproc}, the cooperation information exchange between two transmitters TX~$i$ and TX~$k$ is illustrated.
TX~$k$ sends to TX~$i$ a suitably quantized version of initial CSI estimate $\hat{\mathbf{h}}^{(k)}$, denoted as $\mathbf{z}_{ki}$. A similar operation is performed at TX~$i$ to send $\mathbf{z}_{ik}$ to TX~$k$. The quantization operation associated to the link from TX~$k$ to TX~$i$ is defined as $\mathcal{Q}_{ki}:\mathbb{C}^{n}\mapsto \mathcal{C}_{ki}$, $\mathbf{z}_{ki}\in\mathcal{C}_{ki}$, $\mid\mathcal{C}_{ki}\mid=2^{R_{ki}}$ where $\mathcal{C}_{ki}$ is the codebook for the quantizer $\mathcal{Q}_{ki}$, $n=NMKL$ is the length of the quantization vector.

\subsection{Channel estimation with limited coordination}
\label{ssec:chestlimited}

At TX~$i$, a reconstruction function $g_i(.)$ combines the initial CSI $\hat{\mathbf{h}}^{(i)}$ and the exchanged CSI $\mathbf{z}_{ki}$ to form a final estimate $\tilde{\mathbf{h}}^{(i)}$.

The MMSE estimation problem at TX~$i$ can be formulated as follows:
\begin{align}
D^{(i)}&=\min\frac{1}{n}\mathbb{E}\{\parallel\mathbf{h}-\tilde{\mathbf{h}}^{(i)}\parallel^{2}\}\label{eq:distortion}\\
&=\underset{g_i,\mathcal{Q}_{ki}}\min \frac{1}{n}\mathbb{E}\{\parallel\mathbf{h}-g_i(\hat{\mathbf{h}}^{(i)},\mathcal{Q}_{ki}(\hat{\mathbf{h}}^{(k)}))\parallel^{2}\},\label{generalprob}
\end{align}
where $\tilde{\mathbf{h}}^{(i)}=g_i(\hat{\mathbf{h}}^{(i)},\mathcal{Q}_{ki}(\hat{\mathbf{h}}^{(k)}))$.

Note that it is in general a difficult functional optimization and the two functions $g_i(.)$ and $\mathcal{Q}_{ki}$ are intertwined. The goal of this work is to find (i) a suitable reconstruction function $g_i(.)$ and (ii) the optimal quantizer $\mathcal{Q}_{ki}$ such that at TX~$i$, $D^{(i)}$ is minimized.

\section{Optimal vector quantization model}
\label{sec:quantmod}
We now first introduce a useful model for the optimal vector quantization (VQ) which will be exploited in the latter analysis.

\begin{figure}[tp!]
\centering
\begin{minipage}[t]{0.495\textwidth}
\centering
\includegraphics[width=1.0\columnwidth]{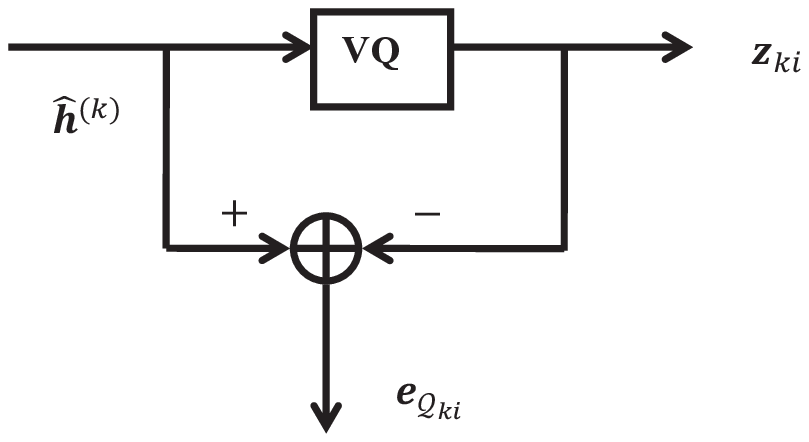}
\caption{Quantizer model for optimal vector quantization.}\label{fig:optVQ}
\end{minipage}
\begin{minipage}[t]{0.495\textwidth}
\centering
\includegraphics[width=1.0\columnwidth]{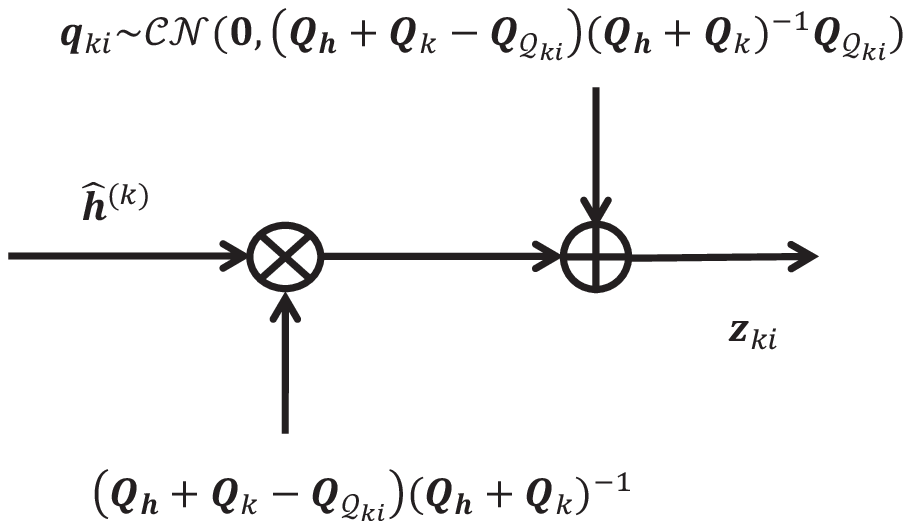}
\caption{Gain-plus-additive-noise model for the optimal vector quantization procedure.}\label{fig:mchmod}
\end{minipage}
\end{figure}

Generally, optimal VQ can be derived via a Lloyd-Max algorithm as depicted in Fig. \ref{fig:optVQ}. The quantization result $\mathbf{z}_{ki}$ and quantization error $\mathbf{e}_{\mathcal{Q}_{ki}}$ are uncorrelated but the quantization input $\hat{\mathbf{h}}^{(k)}$ is both dependent on $\mathbf{e}_{\mathcal{Q}_{ki}}$ and $\mathbf{z}_{ki}$. The covariance matrices for $\hat{\mathbf{h}}^{(k)}$, $\mathbf{z}_{ki}$ and $\mathbf{e}_{\mathcal{Q}_{ki}}$ satisfy \cite{705449}
\begin{align}
\mathbf{Q}_{\hat{\mathbf{h}}^{(k)}}=\mathbf{Q}_{\mathbf{z}_{ki}}+\mathbf{Q}_{\mathcal{Q}_{ki}}.
\end{align}
Since the input of the quantizer $\hat{\mathbf{h}}^{(k)}$ is Gaussian, we obtain an upper bound of the quantization impact by assuming that the quantization error
\begin{align}
\mathbf{e}_{\mathcal{Q}_{ki}}=\hat{\mathbf{h}}^{(k)}-\mathbf{z}_{ki}
\end{align}
is also Gaussian distributed as $\mathbf{e}_{\mathcal{Q}_{ki}}\sim\mathcal{CN}(\mathbf{0},\mathbf{Q}_{\mathcal{Q}_{ki}})$ \cite{cover2012elements}. Similar to \cite{5425382} and based on the Gaussian assumption for $\mathbf{e}_{\mathcal{Q}_{ki}}$, we can approximate the VQ procedure by a gain-plus-additive-noise model (similar to the scalar quantizer case in \cite{westerink1992scalar}) as illustrated in Fig.\ref{fig:mchmod}.
\newtheorem{corollary1}{Proposition}
\begin{corollary1}
\label{coro4}
Assuming that the quantization error $\mathbf{e}_{\mathcal{Q}_{ki}}\sim\mathcal{CN}(\mathbf{0},\mathbf{Q}_{\mathcal{Q}_{ki}})$ and is independent from the quantization result $\mathbf{z}_{ki}$, the optimal vector quantization for $\hat{\mathbf{h}}^{(k)}\sim\mathcal{CN}(\mathbf{0},\mathbf{Q}_{\mathbf{h}}+\mathbf{Q}_{k})$ is given by a gain-plus-additive-noise model:
\begin{equation}\label{zki}
\mathbf{z}_{ki}=(\mathbf{Q}_{\mathbf{h}}+\mathbf{Q}_{k}-\mathbf{Q}_{\mathcal{Q}_{ki}})(\mathbf{Q}_{\mathbf{h}}+\mathbf{Q}_{k})^{-1}\hat{\mathbf{h}}^{(k)}+\mathbf{q}_{ki},
\end{equation}
where $\mathbf{q}_{ki}$ and $\hat{\mathbf{h}}^{(k)}$ are uncorrelated random vectors, $\mathbf{q}_{ki}\sim\mathcal{CN}(\mathbf{0},(\mathbf{Q}_{\mathbf{h}}+\mathbf{Q}_{k}-\mathbf{Q}_{\mathcal{Q}_{ki}})(\mathbf{Q}_{\mathbf{h}}+\mathbf{Q}_{k})^{-1}\mathbf{Q}_{\mathcal{Q}_{ki}})$.
\end{corollary1}
\begin{proof}
Since $\mathbf{e}_{\mathcal{Q}_{ki}}$ is assumed to be independent from $\mathbf{z}_{ki}$. Knowing that $\hat{\mathbf{h}}^{(k)}=\mathbf{e}_{\mathcal{Q}_{ki}}+\mathbf{z}_{ki}$, $\mathbf{e}_{\mathcal{Q}_{ki}}\sim\mathcal{CN}(\mathbf{0},\mathbf{Q}_{\mathcal{Q}_{ki}})$, $\mathbf{h}\sim\mathcal{CN}(\mathbf{0},\mathbf{Q}_{\mathbf{h}})$, according to the Bayesian estimator \cite{kay1993fundamentals},
\begin{equation*}
\begin{array}{lcl}
\mathbb{E}\{\mathbf{z}_{ki}|\hat{\mathbf{h}}^{(k)}\}&=&(\mathbf{Q}_{\mathbf{h}}+\mathbf{Q}_{k}-\mathbf{Q}_{\mathcal{Q}_{ki}})(\mathbf{Q}_{\mathbf{h}}+\mathbf{Q}_{k})^{-1}\hat{\mathbf{h}}^{(k)}\\
\Cov\{\mathbf{z}_{ki}|\hat{\mathbf{h}}^{(k)}\}&=&(\mathbf{Q}_{\mathbf{h}}+\mathbf{Q}_{k}-\mathbf{Q}_{\mathcal{Q}_{ki}})(\mathbf{Q}_{\mathbf{h}}+\mathbf{Q}_{k})^{-1}\mathbf{Q}_{\mathcal{Q}_{ki}},\\
\end{array}
\end{equation*}
which concludes the proof.
\end{proof}

This gain-plus-additive noise model with uncorrelated $\mathbf{z}_{ki}$ and $\mathbf{q}_{ki}$ is helpful in the following derivation. In the reminder of this work, both the design of the reconstruction functions and optimal quantizers will be based on (\ref{zki}).

\section{reconstruction function design}
\label{sec:estcomb}
This section addresses the aforementioned sub-problem of the optimal reconstruction function design in general settings of multi-TXs cooperation ($K\geq2$). For ease of illustration, we focus on TX~$i$, who is cooperating with TX~$k, \forall k\in \mathcal{A}_i$, where the set $\mathcal{A}_i$ contains the indices of TXs that are cooperating with TX~$i$.

We consider hereby the reconstruction function as a weighted linear combination of estimates at TX~$i$, which is suboptimal for the functional optimization in \eqref{generalprob} but leads to a desired closed form optimization. It should be noticed that in the particular case of $K=2$, we have proven in Section \ref{ssec:asymres} that a weighted linear reconstruction function is actually optimal as it achieves asymptotically the Wyner-Ziv bound.

Hence, the final estimate at TX~$i$ is modeled as:
\begin{align}\label{finalest}
%\tilde{\mathbf{h}}^{(i)}=&\underset{k\in\mathcal{A}_i}\sum\mathbf{W}_{ki}\mathbf{z}_{ki}\nonumber\\
%s.t.&\underset{k\in\mathcal{A}_i}\sum\mathbf{W}_{ki}=\mathbf{I},
\tilde{\mathbf{h}}^{(i)}=&\underset{k\in\mathcal{A}_i}\sum\mathbf{W}_{ki}\mathbf{z}_{ki}+\mathbf{W}_{ii}\hat{\mathbf{h}}^{(i)},
\end{align}
where $\mathbf{W}_{ki},\mathbf{W}_{ii}$ are weighting matrices. The optimal weight combining matrices can be obtained in the following proposition.
\newtheorem{corollary2}[corollary1]{Proposition}
\begin{corollary2}\label{coro4}
Let's consider a multi-TXs cooperation described in \eqref{finalest}. Assuming that the CSI estimate at each TX is distributed according to section \ref{ssec:DCSImod} and the limited rate coordination is modeled according to section \ref{ssec:lrcmod}. Let the quantization error covariance matrices be denoted as $\mathbf{Q}_{\mathcal{Q}_{ki}},\forall k\in\mathcal{A}_i$. The optimum per dimensional MSE for the final estimate at TX~$i$ is:
\begin{equation}\label{optDi}
D^{(i)opt}  = \frac{1}{n}\Tr\left(\mathbf{Q}_{\mathbf{h}}^{-1}+\mathbf{Q}_i^{-1}+\bm\Lambda_i\right)^{-1},
\end{equation}
where $\bm\Lambda_i$ is defined as:
\begin{align*}
\bm\Lambda_i\!=\!\underset{k\in\mathcal{A}_i}{\sum}\!\left((\mathbf{Q}_{\mathbf{h}}\!+\!\mathbf{Q}_k)\!\left(\mathbf{Q}_{\mathbf{h}}\!+\!\mathbf{Q}_{k}\!-\!\mathbf{Q}_{\mathcal{Q}_{k}}\!\right)^{-1}(\mathbf{Q}_{\mathbf{h}}\!+\!\mathbf{Q}_k)\!-\!\mathbf{Q}_{\mathbf{h}}\!\right)^{-1}.
\end{align*}
The optimal weight combing matrices $\{\mathbf{W}_{ki}^{opt},\mathbf{W}_{ii}^{opt},\forall k\in\mathcal{A}_i\}$
are obtained as:
\begin{equation}\label{optW}
\left[\begin{array}{cccc}
\mathbf{W}_{ii}^{opt} & \mathbf{W}_{l_1i}^{opt} & \ldots & \mathbf{W}_{l_{C_i}i}^{opt}
\end{array}\right]
=\mathbf{Q}_{\mathbf{h}}\bm\Upsilon_i\bm\Omega_i^{-1},
\end{equation}
where $\bm\Upsilon_i, \bm\Omega_i$ are given below, the set $\mathcal{A}_i$ has cardinality $|\mathcal{A}_i|=C_i$ and each element in the set is denoted by $\mathcal{A}_i=\{l_1,\dots,l_{C_i}\}$.
\begin{equation*}
\begin{array}{lcl}
\mathbf{P}_{l_ji}\! &\! =\! &\! \mathbf{Q}_{\mathbf{h}}+\mathbf{Q}_{l_j}-\mathbf{Q}_{\mathcal{Q}_{l_ji}}\\
\mathbf{A}_{l_ji}\! &\! =\! &\! \mathbf{P}_{l_ji}(\mathbf{Q}_{\mathbf{h}}+\mathbf{Q}_{l_j})^{-1},\qquad \forall j=1\dots C_i\\
\bm\Upsilon_i\! & \!=\! & \!\left[\begin{array}{cccc}
\mathbf{I} & \mathbf{A}_{l_1i}^{H} & \ldots & \mathbf{A}_{l_{C_i}i}^{H}
\end{array}\right]\\
\bm\Omega_i & =& \left[\begin{array}{cccc}
\mathbf{Q}_{\mathbf{h}}+\mathbf{Q}_i & \mathbf{Q}_{\mathbf{h}}\mathbf{A}_{l_1i}^{H} & \ldots & \mathbf{Q}_{\mathbf{h}}\mathbf{A}_{l_{C_i}i}^{H}\\
\mathbf{A}_{l_1i}\mathbf{Q}_{\mathbf{h}} & \mathbf{P}_{l_1i} & \ldots & \mathbf{A}_{l_1i}\mathbf{Q}_{\mathbf{h}}\mathbf{A}_{l_{C_i}i}^{H}\\
\vdots & \vdots & \ddots & \vdots\\
\mathbf{A}_{l_{C_i}i}\mathbf{Q}_{\mathbf{h}} & \mathbf{A}_{l_{C_i}i}\mathbf{Q}_{\mathbf{h}}\mathbf{A}_{l_1i}^H & \ldots & \mathbf{P}_{l_{C_i}i}
\end{array}\right].\\
\end{array}
\end{equation*}
\end{corollary2}
\begin{proof}
See Appendix \ref{sec:appenA}.
\end{proof}
\theoremstyle{remark}\newtheorem{remark1}{Remark}
\begin{remark1}
The optimal weight combining matrices $\{\mathbf{W}_{ki}^{opt},\mathbf{W}_{ii}^{opt},\forall k\in\mathcal{A}_i\}$ and $D^{(i)opt}$ are merely functions of statistics $\mathbf{Q}_{\mathbf{h}},\mathbf{Q}_i,\mathbf{Q}_k,\mathbf{Q}_{\mathcal{Q}_{ki}},\forall k\in\mathcal{A}_i$.\qed
\end{remark1}

\theoremstyle{remark}\newtheorem{remark2}[remark1]{Remark}
\begin{remark2}
Consider a motivation example of two TXs cooperation, at TX~$1$, the final estimate is:
\begin{equation*}
\tilde{\mathbf{h}}^{(1)}=\mathbf{W}_{21}\mathbf{z}_{21}+\mathbf{W}_{11}\hat{\mathbf{h}}^{(1)},
\end{equation*}
where
\begin{equation*}
\begin{array}{lcl}
\mathbf{P}_{21} & = & \mathbf{Q}_{\mathbf{h}}+\mathbf{Q}_2-\mathbf{Q}_{\mathcal{Q}_{21}}\\
\mathbf{A}_{21} & = & \mathbf{P}_{21}(\mathbf{Q}_{\mathbf{h}}+\mathbf{Q}_2)^{-1}\\
\left[\mathbf{W}_{11},\mathbf{W}_{21}\right] & = & \mathbf{Q}_{\mathbf{h}}\left[\begin{array}{ll}\mathbf{I}&\mathbf{A}_{21}^H\end{array}\right]\left[\begin{array}{ll}\mathbf{Q}_{\mathbf{h}}+\mathbf{Q}_1 & \mathbf{Q}_{\mathbf{h}}\mathbf{A}_{21}^H\\ \mathbf{A}_{21}\mathbf{Q}_{\mathbf{h}} & \mathbf{P}_{21}\\ \end{array}\right]^{-1}.
\end{array}
\end{equation*}
The optimal per dimensional MSE is:
\begin{equation*}
\begin{array}{lcl}
D^{(1)opt} & = & \frac{1}{n}\Tr\left(\mathbf{Q}_{\mathbf{h}}^{-1}+\mathbf{Q}_1^{-1}+\bm\Lambda_1\right)^{-1}\\
\bm\Lambda_1 & = & \left((\mathbf{Q}_{\mathbf{h}}+\mathbf{Q}_2)\mathbf{P}_{21}^{-1}(\mathbf{Q}_{\mathbf{h}}+\mathbf{Q}_2)-\mathbf{Q}_{\mathbf{h}}\right)^{-1}.\\
\end{array}
\end{equation*}
\qed
\end{remark2}

\theoremstyle{remark}\newtheorem{remark3}[remark1]{Remark}
\begin{remark3}
The optimal per dimensional MSE and the error covariance matrix for the final estimate is related to $3$ covariance terms: $\mathbf{Q}_{\mathbf{h}}$ indicates the intrinsic (true) channel statistics, $\mathbf{Q}_i$ refers to the initial estimation error covariance and $\bm\Lambda_i$ is related to the initial estimation error and the quantization error covariance at all TXs that cooperate with TX$~i$. The covariance of the final estimate is formulated as the inverse of the sum of the $3$ terms' individual inverse.\qed
\end{remark3}

\section{Quantizer design}
\label{sec:qzerd}

For the optimal quantizer design, it should be noticed that a conventional optimal VQ (optimal VQ with MSE distortion) implemented by Lloyd-Max algorithm is far from being optimal because rather than minimizing the per dimensional MSE $D^{(i)}$ for the final estimate, it only guarantees that the quantization distortion will be minimized.

Therefore, the quantizer should be properly shaped such that the quantization procedure ensures not only the minimization of quantization distortion, but also guarantees that the quantization result, after weighted combination with other estimates, will have the minimal per dimensional MSE $D^{(i)}$. A useful interpretation of this approach is as follows. As $\mathbf{Q}_{i}, \forall i=1\dots K$ reflect the spatial distribution of accuracy of the initial CSI, the quantizer $\mathcal{Q}_{ki}$ should allocate the quantization resource where more bits are needed, i.e., in channel elements or directions that are well known by TX~$k$ and least known by TX~$i$. To this end, we choose the weighted square error distortion
\begin{align}
d_{\mathcal{Q}_{ki}}(\mathbf{x},\mathbf{y})=(\mathbf{x}-\mathbf{y})^H\mathbf{B}_{ki}(\mathbf{x}-\mathbf{y})
\end{align}
as the distortion measure of the quantizer $\mathcal{Q}_{ki}$ with the positive definite shaping matrix $\mathbf{B}_{ki}$ to be optimized.

For an important intermediate step, we can calculate $\mathbf{Q}_{\mathcal{Q}_{ki}}$ in the asymptotic case as a function of $\mathbf{B}_{ki}$ when the given coordination link rate $R_{ki}$ is sufficiently large, i.e. in the high resolution regime.
\theoremstyle{plain}\newtheorem{corollary3}[corollary1]{Proposition}
\begin{corollary3}
\label{coro1}
Consider the quantization of a $n$ dimensional complex random vector source $\mathbf{x}\sim\mathcal{CN}(\mathbf{0},\bm\Gamma)$, in the high resolution regime where the number of quantization levels $S$ is large, the optimal VQ using weighted MSE distortion with shaping matrix $\mathbf{B}$ will have a quantization error covariance matrix $\mathbf{Q}_{\mathbf{x}}$ as:
\begin{equation*}
\mathbf{Q}_{\mathbf{x}}=\mathbf{Q}_0^{(S)}(\bm\Gamma)\det(\mathbf{B})^{\frac{1}{n}}\mathbf{B}^{-1},
\end{equation*}
where
\begin{equation*}
\mathbf{Q}_{0}^{(S)}(\bm\Gamma)=S^{-\frac{1}{n}}M_{2n}2\pi\left(\frac{n+1}{n}\right)^{n+1}\det(\bm\Gamma)^{\frac{1}{n}}\mathbf{I}_n
\end{equation*}
is the quantization error covariance matrix for $\mathbf{x}$ in the high resolution regime when conventional optimal VQ is applied \cite{1056490}. $M_{2n}$ is a constant related to $2n$.
\end{corollary3}
\begin{proof}
See Appendix \ref{sec:appenB}.
\end{proof}

\theoremstyle{remark}\newtheorem{remark4}[remark1]{Remark}
\begin{remark4}\label{rmk4}
The aforementioned quantization error covariance matrix expression encompasses the quantization error covariance matrix for conventional optimal VQ by taking $\mathbf{B}=\mathbf{I}_n$. It can be easily verified that by imposing a constraint that $\det(\mathbf{B})=1$, for all values of matrix $\mathbf{B}$, the corresponding quantizers will have the same quantization distortion. \qed
\end{remark4}
\theoremstyle{remark}\newtheorem{remark5}[remark1]{Remark}
\begin{remark5}
For the constant $M_{2n}$, a look-up table for $2n=1,\dots,10$ in \cite{1056483} can be used. when $n$ is larger, we can approximate $M_{2n}=\frac{1}{2\pi{e}}$.\qed
\end{remark5}

We now exploit Proposition \ref{coro1} in order to derive $\mathbf{Q}_{\mathcal{Q}_{ki}}$:
\begin{equation}\label{Qquantki}
\mathbf{Q}_{\mathcal{Q}_{ki}}=\mathbf{Q}_{0}^{(S)}(\bm\Gamma)\det(\mathbf{B}_{ki})^{\frac{1}{n}}\mathbf{B}_{ki}^{-1},
\end{equation}
where
\begin{equation*}
\begin{array}{lcl}
S&=&2^{R_{ki}}\\
\bm\Gamma&=&\mathbf{Q}_{\mathbf{h}}+\mathbf{Q}_k.\\
\end{array}
\end{equation*}

\subsection{Shaping matrix optimization}
\label{ssec:optforB}

Based on the reconstruction function in Section \ref{sec:estcomb} and using equations (\ref{optDi}), (\ref{Qquantki}), we can now proceed with the task of jointly optimizing the reconstruction function and the quantizer by solely optimizing the value of $\mathbf{B}_{ki}$.
\begin{equation}\label{optimizeB}
\begin{array}{cl}
\underset{\mathbf{B}_{ki},k\in\mathcal{A}_i}{\min} & D^{(i)opt}\\
s.t. &\det(\mathbf{B}_{ki})=1,\mathbf{B}_{ki}\succeq 0\\
 &D^{(i)opt}  \text{ defined in (\ref{optDi})}.\\
\end{array}
\end{equation}
As mentioned in Remark \ref{rmk4}, the constraints on $\mathbf{B}_{ki}$ matrices ensure that all feasible quantizers have the same quantization distortion, the optimization will find $\mathbf{B}_{ki},\forall k\in\mathcal{A}_i$ that minimize the per dimensional MSE for the final estimate.

\newtheorem{corollary5}[corollary1]{Proposition}
\begin{corollary5}\label{coro5}
The objective function in problem (\ref{optimizeB}) is convex. In high resolution regime, problem (\ref{optimizeB}) can be approximated by the following convex optimization problem:
\begin{equation}\label{optimizeC}
\begin{array}{cl}
\underset{\mathbf{B}_{ki},k\in\mathcal{A}_i}{\min} & \frac{1}{n}\!\Tr\!\left(\!\underset{k\in\mathcal{A}_i}{\sum}\!(\!\mathbf{Q}_k^{-1}\!-\!\mathbf{Q}_k^{-1}\mathbf{Q}_{\mathcal{Q}_{ki}}\mathbf{Q}_k^{-1}\!)\!+\!\mathbf{Q}_i^{-1}\!+\!\mathbf{Q}_{\mathbf{h}}^{-1}\!\right)^{-1}\\
s.t. &\det(\mathbf{B}_{ki})\geq1,\mathbf{B}_{ki}\succeq 0\\
 &\mathbf{Q}_{\mathcal{Q}_{ki}}  \text{ defined in (\ref{Qquantki})}.\\
\end{array}
\end{equation}
\end{corollary5}
\begin{proof}
See Appendix \ref{sec:appenC}.
\end{proof}
The reason for solving optimization problem (\ref{optimizeC}) rather than solving directly the original optimization problem (\ref{optimizeB}) is that the former can be easily transformed into a semi-definite quadratic linear programming. It can be solved efficiently by optimization toolbox such as CVX.

\subsection{Asymptotic result for two TXs cooperation}
\label{ssec:asymres}
Interestingly, the asymptotic performance of the proposed algorithm can be characterized in relation to known information theoretic bound.
\newtheorem{corollary6}[corollary1]{Proposition}
\begin{corollary6}\label{coro6}
For a two TXs cooperation scenario of TX~$1$ and TX~$2$ as described in section \ref{ssec:chestlimited}, at TX~$1$ the proposed coordination shaping algorithm can achieve asymptotically in high resolution regime the Wyner-Ziv bound given by
\begin{align*}
D^{(1)opt}_{\infty}=D^{(1)NWZ}_{\infty}=\frac{1}{n}\Tr\left(\mathbf{Q}_1^{-1}+\mathbf{Q}_{2}^{-1}+\mathbf{Q}_{\mathbf{h}}^{-1}\right)^{-1}.
\end{align*}
\end{corollary6}
\begin{proof}
the asymptotical per dimension MSE for proposed algorithm is:
\begin{align*}
D^{(1)opt}_{\infty}&=\lim_{R_{21}\rightarrow\infty}D^{(1)opt}\\
&=\frac{1}{n}\Tr\left(\mathbf{Q}_1^{-1}+\mathbf{Q}_{2}^{-1}+\mathbf{Q}_{\mathbf{h}}^{-1}\right)^{-1}.
\end{align*}
It is well known that the information theoretic bound of the per dimension MSE for two TXs cooperation can be achieved using a Wyner-Ziv quantizer and the asymptotic distortion is \cite{rebollo2004wyner}:
\begin{equation*}
D^{(1)NWZ}_{\infty}=\frac{1}{n}\Tr(\mathbb{E}_{YZ}\Var{[X|Y,Z]}),
\end{equation*}
where X, Y, Z correspond to the source data, side information and noisy source (i.e, perfect CSI $\mathbf{h}$, initial CSI $\hat{\mathbf{h}}^{(1)}$ and initial CSI at it's cooperation TX $\hat{\mathbf{h}}^{(2)}$ as in our case).
Since Gaussianity is assumed for the perfect CSI and initial CSI, $D^{(1)NWZ}_{\infty}$ can be calculated as:
\begin{align*}
&D^{(1)NWZ}_{\infty}\\
=&\frac{1}{n}\!\Tr\!\left(\!\mathbf{Q}_{\mathbf{h}}\!-\!\left[\!\begin{array}{cc}\mathbf{Q}_{\mathbf{h}}&\mathbf{Q}_{\mathbf{h}}\end{array}\!\right]\!
\left[\!\begin{array}{cc}\mathbf{Q}_{\mathbf{h}}\!+\!\mathbf{Q_1}&\mathbf{Q}_{\mathbf{h}}\\
\mathbf{Q}_{\mathbf{h}}&\mathbf{Q}_{\mathbf{h}}\!+\!\mathbf{Q_{2}}
\end{array}\!\right]^{-1}\!\left[\!\begin{array}{c}\mathbf{Q}_{\mathbf{h}}\\ \mathbf{Q}_{\mathbf{h}}\end{array}\!\right]\!\right)\\
=&\frac{1}{n}\Tr\left(\mathbf{T}_1-\mathbf{T}_1(\mathbf{T}_1+\mathbf{Q}_{2})^{-1}\mathbf{T}_1\right)\\
=&\frac{1}{n}\Tr\left(\mathbf{T}_1^{-1}+\mathbf{Q}_{2}^{-1}\right)^{-1}\\
=&\frac{1}{n}\Tr\left(\mathbf{Q}_1^{-1}+\mathbf{Q}_{2}^{-1}+\mathbf{Q}_{\mathbf{h}}^{-1}\right)^{-1}\\
=&D^{(1)opt}_{\infty},
\end{align*}
where $\mathbf{T}_1=\mathbf{Q}_{\mathbf{h}}(\mathbf{Q}_{\mathbf{h}}+\mathbf{Q}_1)^{-1}\mathbf{Q}_{1}$.
\end{proof}
Thus, it reveals that in the case of two TXs cooperation, the proposed coordination shaping algorithm is asymptotically optimal.

\subsection{The shaped coordination design}
\label{ssec:shapedcoorddesign}
\begin{figure}[tp!]
\centering
\includegraphics[width=1\columnwidth]{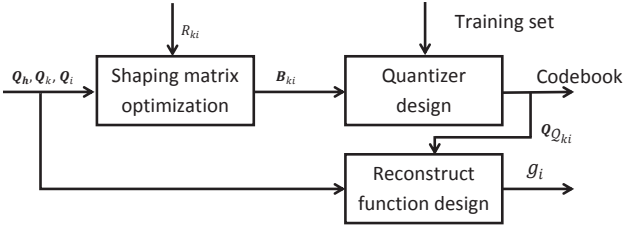}
\caption{The procedure of optimal shaped coordination design.
}\label{fig:shapedcoorddesign}
\end{figure}

The shaped coordination design is depicted in Fig. \ref{fig:shapedcoorddesign}.
During the quantizer design, each TX will first optimize the shaping matrices based on channel statistics and the number of bits for information exchange on each backhaul. Once the optimal shaping matrix is obtained, the codebook for the optimal quantizer can be calculated based on the Lloyd algorithm and a training set. The optimal weight matrices for the reconstruct function can be obtained according to (\ref{optW}). It should be noticed that this shaped coordination quantizer and the reconstruction function design is semi-static. The weight matrices for the estimates combination and the quantizers will be updated only when the channel statistics or the backhaul resources allocation have been changed.

After the shaped coordination design, the optimal instantaneous final channel estimate at each TX can be attained based on the shaped coordination procedure in Fig. \ref{fig:shapedcoordproc}.

\section{Complexity analysis for shaped coordination}
\label{sec:companalysis}
In this section, we will compare the complexity of the proposed shaped coordination algorithm with state of the art Wyner-Ziv coding implementations in two aspects. In the first part, we will compare the complexity of the quantizer and the reconstruction function design for the proposed algorithm with state of the art design. In the second part, we will compare the complexity to obtain final channel estimate using the proposed shaped coordination algorithm with state of the art Wyner-Ziv coding implementation.

\subsection{Complexity analysis for quantizer and reconstruct function design}
As described in Fig. \ref{fig:shapedcoorddesign}, the quantizer design for the proposed algorithm involves a semi-definite program to find the optimal shaping matrix and a lloyd max procedure to acquire the codeword for the quantizer.

Solving a semi-definite program using the ellipsoid method requires an overall complexity of $O(\max(m,n^2)n^6\log(\frac{1}{\epsilon}))$ \cite{bubeck2014convex}, where $n$ is the dimension of optimization matrix, $m$ is the number of inequality constraints, $\epsilon$ is the accuracy required for the iteration process. In a two TX cooperation case, the overall complexity to find an $\epsilon$-optimal shaping matrix is $O(n^8\log(\frac{1}{\epsilon}))$ where $n$ is the dimension of the shaping matrix. Regarding to the lloyd max algorithm, the complexity is often given as $O(\ell SCP)$, where $S$ is the size of the training set, $C=2^R$ is the number of clusters, $P$ is the number of iteration needed until convergence, $\ell=2n$ is the dimension of the training vector. In practice, the convergence of iteration is fast and the overall algorithm has polynomial smoothed running time \cite{arthur2009k}. Therefore, the overall complexity of the proposed shaped coordination design is $O(n^8\log(\frac{1}{\epsilon})+2^{R+1}nSP)$. In real implementation, the complexity mainly comes from lloyd max codebook design.

For the state of the art design of noisy source Wyner-Ziv coding problem, possible implementation can be nested lattice quantizer \cite{liu2006slepian} or Wyner-Ziv transform coding \cite{rebollo2004wyner}. For the optimal nested lattice design, it involves the design of lattices with the densest packing (i.e., the best channel code) and the thinnest covering (i.e., the best source coding) \cite{liu2006slepian}. In addition, in order to optimize the nesting scheme, the densest packing lattice should have Voronoi cell boundaries not intersecting with the thinnest covering lattice (this property is referred as clean in \cite{conway2002existence}). It should be guaranteed that the two lattices are geometrically similar and the nesting ratio should be $2^R$ \cite{liu2006slepian}. For certain small dimensions, the thinnest covering lattice is already known \cite{1056483} and it is still an open problem for arbitrary dimensions. However, as the lloyd max algorithm is reported to converge to the centroidal Voronoi tessellation \cite{du2006convergence}, the complexity of finding the thinnest covering is \emph{at least as hard as the lloyd max algorithm}. Regarding to the problem of finding a clean and similar densest packing lattice which also satisfies the nesting ratio $2^R$, to the best of the authors' knowledge, it's a difficult open problem and only partial result is given in \cite{conway2002existence}. Therefore, the complexity of nested quantizer design is higher than the proposed shaped coordination design.

Consider the Wyner-Ziv transform coding design, the overall system requires $2n$ parallel path of scalar quantization followed by ideal Slepian-Wolf coding and decoding \cite{rebollo2004wyner}. The Slepian-Wolf coding and decoding can be implemented using a rate compatible punctured turbo code \cite{Aaron2002compression} or polar code. The quantizer design using lloyd max algorithm for each path has a complexity of $O(2^{\frac{R}{2n}}SP')$, where $S$ is the size of the training set, $2^{\frac{R}{2n}}$ is the number of clusters for each path, $P'$ is the number of iteration needed until convergence. The overall complexity for the $2n$ path is $O(n2^{\frac{R}{2n}+1}SP')$. In real implementation, the running time complexity of the lloyd max codebook design is dominated by the cardinality $S$ of the training set. Therefore, the overall complexity for the proposed shaped coordination design is comparable to the Wyner-Ziv transform coding design.

\subsection{Complexity analysis for obtaining final CSI estimate}
Fig. \ref{fig:shapedcoordproc} has introduced the procedure to obtain final CSI estimate using shaped coordination algorithm. It can be concluded that there will be $1$ vector quantization and $1$ weighted combination so as to get the final CSI estimate for $2$ TXs cooperation. By using fast quantization algorithm such as vector to scalar mapping elimination \cite{lin1996dynamic}, the complexity of the vector quantization is $O(10n+R+6nN_M-3)$, where $N_M$ is the average number of affected code vectors. The weighted combination has a complexity of $O(8n^2+2n)$. Therefore, the overall complexity is $O(8n^2+12n+R+6nN_M-3)$.

To obtain the final CSI estimate using Wyner-Ziv transform coding, the complexity for the scalar quantization in each path is $O(\frac{R}{2n})$ when a binary search algorithm is applied. The complexity for the rate compatible punctured turbo code is hard to obtain. However, For an equivalent polar code implementation of the ideal Slepian-Wolf coding and decoding, the encoding or decoding complexity is well known as $O(N\log{N})$ where $N=\frac{R}{2n}$. Therefore, the overall complexity for Wyner-Ziv transform coding is $O(8n^2+2n+R+2R\log{\frac{R}{2n}})$, where the $8n^2+2n$ term comes from the pre-processing and post-processing of orthogonal transformation and the final combination step.

It is therefore evident that in high resolution regime, the proposed algorithm has a complexity of the order $O(R)$ while the Wyner-Ziv transform coding algorithm has a complexity of $O(R\log{R})$, which is higher than the algorithm we proposed.

\section{Coordination link bit allocation}
\label{sec:coordlinkbitalloc}
An interesting consequence of the above analysis is the optimization of coordination where multiple transmitters can exchange simultaneously CSI-related information to each other under a global constraint on the coordination bits. The optimization problem over all coordination links now becomes:
\begin{equation}\label{cooralloc}
\begin{array}{cl}
\underset{\substack{\mathbf{B}_{ki},R_{ki}\\k\in\mathcal{A}_i,i=1,\dots,K}}{\min} &\frac{1}{K}\underset{i=1}{\overset{K}{\sum}} D^{(i)opt}\\
s.t. &\det(\mathbf{B}_{ki})=1,\mathbf{B}_{ki}\succeq 0\\
&\overset{K}{\underset{i=1}\sum}\underset{k\in\mathcal{A}_i}\sum{R_{ki}}=R_{tot},R_{ki}\in\mathbb{N}^{+}\\
&D^{(i)opt} \text{ defined in (\ref{optDi})}.
\end{array}
\end{equation}

Due to the integer constraints on $R_{ki}$, this problem becomes a non-convex optimization. However, conventional alternating algorithms can be applied to perform the optimization. In this two-step alternative optimization, either the capacity constraints on the backhaul $R_{ki}$ or the shaping matrices $\mathbf{B}_{ki}$ will be fixed while optimizing the other.

The optimization for $R_{ki}$ with fixed $\mathbf{B}_{ki}$ is an integer programming problem and the optimization for $\mathbf{B}_{ki}$ with fixed $R_{ki}$ is a convex optimization problem. Hence, many conventional algorithms can be applied in both steps. It should be noted that the alternating algorithm does not guarantee the global optimum. Based on the initial point, it might converge to a local optimal point as well.

\section{Numerical performance analysis}
\label{sec:nuperf}

\begin{figure}[tp!]
\centering
\includegraphics[width=1\columnwidth]{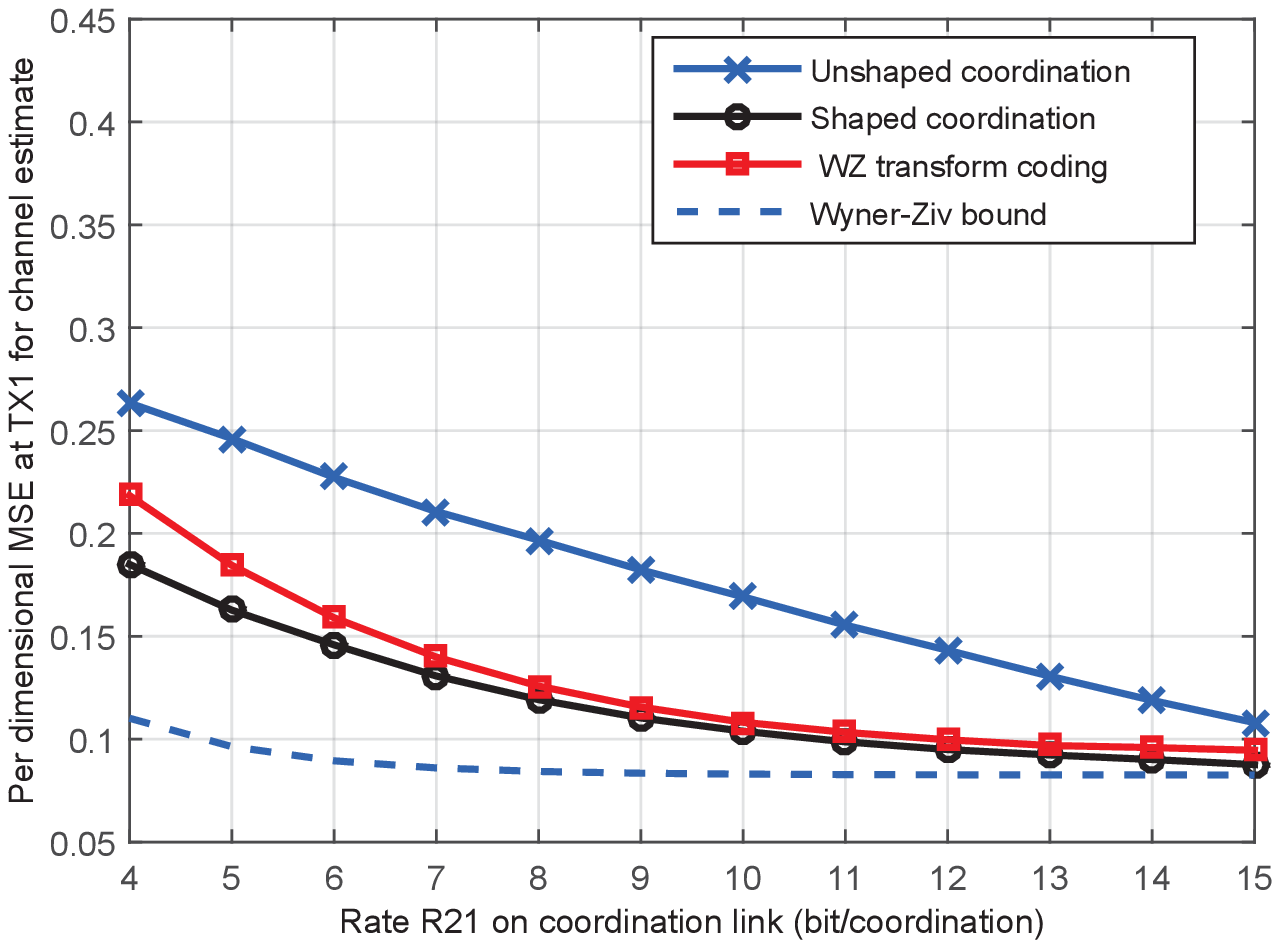}
\includegraphics[width=1\columnwidth]{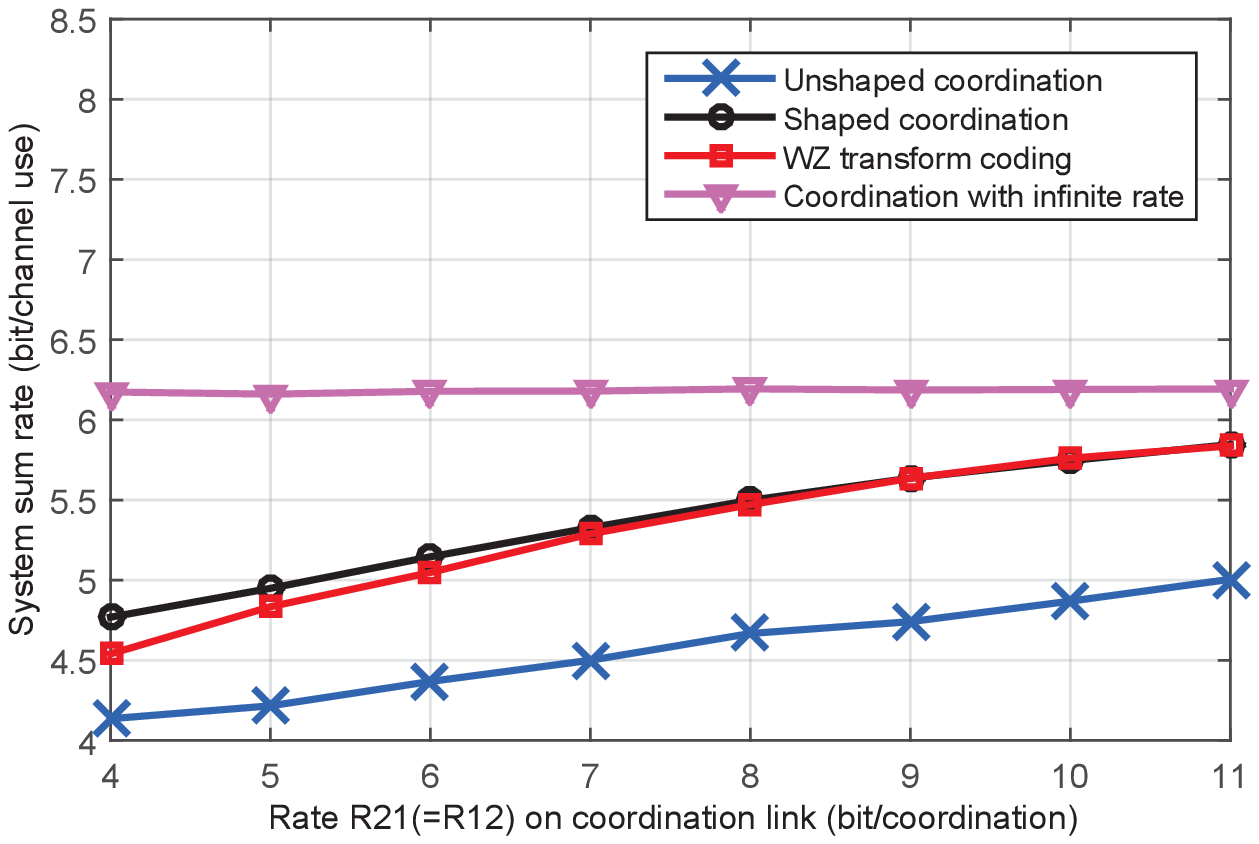}
\caption{MSE for the final channel estimate at TX~$1$ and sum rate for the $2$ TX cooperation system vs. coordination link rate $R_{21}$, CSI error covariance matrices are $\mathbf{Q}_1=diag(0.1,0.9,0.1,0.9)$, $\mathbf{Q}_2=diag(0.9,0.1,0.9,0.1)$, each TX implements a ZF precoder based on its final channel estimation, SNR is $20$dB per TX.
}\label{fig:shapedcoor}
\end{figure}

In this section, a network MIMO transmission \cite{5594708} setup is considered.  Unless otherwise indicated, the default simulation settings are $K=2$ and $L=2$, $M=N=1$. An isotropic channel $\mathbf{h}\in\mathbb{C}^{4\times1}\sim\mathcal{CN}(\mathbf{0},\mathbf{I}_4)$ is considered. The rates on coordination link from TX~$2$ to TX~$1$ and from TX~$1$ to TX~$2$ are denoted $R_{21}$ and $R_{12}$, respectively. Each TX constructs a ZF precoder based on its final channel estimate. The power control at each TX is $20$dB. The per dimensional MSE for decentralized channel estimation is evaluated for different settings using Monte-Carlo simulations over $10^5$ channel realizations.
Since $n=MNKL=4$, the parameter $M_{2n}$ is chosen to be $929/12960$ which is related to the $E8$ lattice\cite{1056483}. In Fig. \ref{fig:shapedcoor}, the CSI information structure is characterized by $\mathbf{Q}_1=diag(0.1,0.9,0.1,0.9)$ and $\mathbf{Q}_2=diag(0.9,0.1,0.9,0.1)$ which corresponds to a broadcast feedback example where TX~$1$ has more accurate CSI on RX~$1$ and less accurate CSI on RX~$2$, and vice versa for TX~$2$. The $diag(.)$ operator represents a diagonal matrix with diagonal elements in the parenthesis.

The first sub-figure in Fig.\ref{fig:shapedcoor} shows the per dimensional MSE for the final estimation at TX~$1$. The Wyner-Ziv bound is the information theoretic bound. The shaped coordination curve applies the proposed algorithm. The unshaped coordination implements the traditional optimal VQ and finds $\mathbf{W}_{21}, \mathbf{W}_{11}$ accordingly using (\ref{optW}). From the figure we can conclude that the shaped coordination algorithm outperforms the unshaped coordination algorithm, which not surprisingly shows the benefit of taking the priori statistic information into account. The WZ transform coding curve refers to the asymptotic optimal Wyner-Ziv transform coding for noisy source in \cite{rebollo2004wyner}. It reveals that our algorithm outperforms the Wyner-Ziv transform coding algorithm in low coordination rate region and converges asymptotically to the Wyner-Ziv bound $D_{\infty}$ as expected. The second sub-figure of Fig.\ref{fig:shapedcoor} exhibits the sum rate for a $2$ TX cooperation system. The rate on the coordination links satisfies $R_{21}=R_{12}$. We also provide the sum rate for the case when coordination links have infinite bandwidth as a baseline. The figure shows that the proposed shaped coordination algorithm will improve the system sum rate beyond the unshaped coordination algorithm and WZ transform coding algorithm when a simple ZF precoder is implemented. As the rate on coordination link increases, the sum rate for all algorithms will converge to the infinite coordination rate case.

\begin{figure}[tp!]
\centering
\includegraphics[width=1\columnwidth]{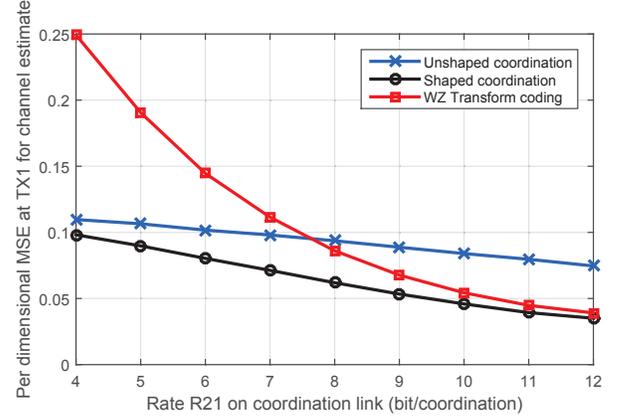}
\caption{MSE for the final channel estimation at TX~$1$ as a function of coordination link rate $R_{21}$ for a $2$ TX $2$ RX cellular simulation.}\label{fig:cellularsimu}
\end{figure}

In Fig.\ref{fig:cellularsimu}, the per dimensional MSE for final CSI estimate at TX~$1$ is plotted as a function of backhaul rate for a cooperation scenario with $K=2$ TXs and $L=2$ RXs. Each RX has single antenna ($N=1$) and each TX has $2$ transmit antennas ($M=2$). we simulate with a realistic cellular setting. Let $\mathbf{h}_{\ell}$ denote the user channel for RX~$\ell$, assuming that $\mathbf{h}_{\ell}\sim\mathcal{CN}(\mathbf{0},\bm\Theta_{\ell})$. The correlation matrix $\bm\Theta_{\ell}$ is a block diagonal matrix that reads  $\bm\Theta_{\ell}=\blockdiag(\bm\Theta_{\ell,1},\ldots,\bm\Theta_{\ell,k})$. The correlation matrix  $\bm\Theta_{\ell,k}$ between the $\ell$th RX and the $k$th TX denotes \cite{wagner2012}
\begin{align}\label{passlossmodeltheta}
[\bm\Theta_{\ell,k}]_{i,j}\!=\!\gamma d_{\ell,k}^{-\epsilon}\cdot\frac{1}{\theta_{\ell,\max}\!-\!\theta_{\ell,\min}}\!\int_{\theta_{\ell,\min}}^{\theta_{\ell,\max}} e^{\mathfrak{i}\frac{2\pi}{\lambda}\cdot (j-i) d_{as}\cdot \cos{\theta}}d\theta.
\end{align}
The $\gamma d_{\ell,k}^{-\epsilon}$ part indicates the pathloss with $\epsilon$ being the pathloss component. $d_{\ell,k}$ the distance between RX~$\ell$ and TX~$k$ and $\gamma$ a coefficient to further adjust the model. The rest part models the Uniform Linear Array (ULA) assuming a diffuse two-dimensional field of isotropic scatters around the receivers. The waves impinge the receiver $\ell$ uniformly at an azimuth angle $\theta$ ranging from $\theta_{\ell,\min}$ to $\theta_{\ell,\max}$. The angle spread is $\varphi=\theta_{\ell,\max}-\theta_{\ell,\min}$, the antennas spacing is $d_{as}$ and $\lambda$ is the signal wavelength. Each TX is located in the center of a hexagon cell with cell radius $r_{c}$, each cell has only $1$ RX and the RX is random and uniformly located in the cell with the distance to the TX larger than $d_0$. Let $\mathbf{e}^{(k)}_{\ell}$ denote the estimate error seen at TX~$k$ for RX~$\ell$ user channel, $\mathbf{e}^{(k)}_{\ell}\sim\mathcal{CN}(\mathbf{0},2^{-r^{\textrm{FB}}_{k\ell}}\cdot\sigma_{E}^2\mathbf{I}).$ The parameter $\sigma_{E}^2=\frac{\sigma_{\textrm{pilot}}^2}{N_{\textrm{pilot}}\cdot p_{\textrm{pilot}}}$ is based on the Cramer-Rao lower bound \cite{kay1993fundamentals}, yielding the absolute mean square error of a channel estimation based on $N_{\textrm{pilot}}$ pilots  of power $p_{\textrm{pilot}}$ subjects to noise $\sigma_{\textrm{pilot}}^2$ \cite{Marsch2011uplinkcomp}. The term $2^{-r^{\textrm{FB}}_{k\ell}}$ is based on rate distortion theory where $r^{\textrm{FB}}_{k\ell}$ is the average rate that per channel coefficient is instantaneously fed back from RX~$\ell$ to TX~$k$.

All simulation parameters are listed in the Table \ref{tablecellular}. This setting corresponds to a scenario of FDD downlink channel estimation and each RX will feedback to both TXs. However, RX feeds back the CSIR to TX in the cell with a higher feedback rate than to the TX in the other cell.
\begin{table}[h]
\begin{center}
\caption{Simulation parameters for a cellular setting.}\label{tablecellular}
\begin{tabular}{|c|c|c|c|c|c|c|c|c|}
\hline
$r_c$ & $\gamma$ & $\epsilon$ & $d_0$ & $f$ & $d_{as}$ & $\varphi$ & $\sigma_{E}^2$ & $r^{\textrm{FB}}_{k\ell}$ \\ \hline
$1km$ & $10^{9}$ & $3$ & $0.1km$ & $2$GHz & $\frac{\lambda}{2}$ & $\frac{\pi}{6}$  & $1$ & \begin{tabular}{c}$5(\ell=k)$\\$1(\ell\neq k)$\end{tabular} \\ \hline
\end{tabular}
\end{center}
\end{table}

From Fig. \ref{fig:cellularsimu}, we can conclude that the proposed shaped coordination algorithm outperforms the Wyner-Ziv transform coding algorithm in low coordination rate region. The Wyner-Ziv transform coding algorithm is even worse than the unshaped coordination probably because in this setting, using an additive separable linear estimator to replace the non-additive separable $\mathbb{E}_{YZ}\Var{[X|Y,Z]}$ is highly suboptimal \cite{rebollo2004wyner}.

\begin{figure}[tp!]
\centering
\includegraphics[width=1\columnwidth]{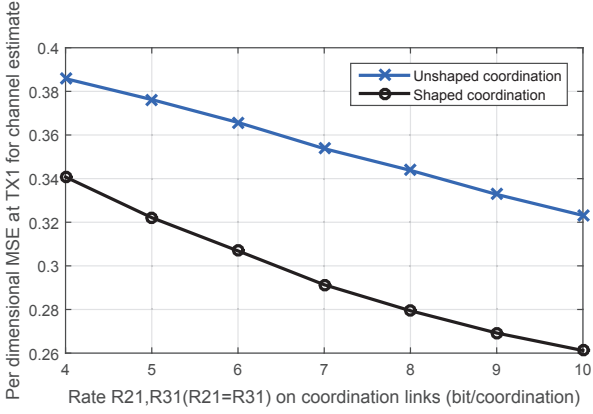}
\caption{$3$ TX cooperation: MSE for the final channel estimate at TX~$1$ as a function of coordination link rate $R_{21}(=R_{31})$.}\label{fig:3TXcoop}
\end{figure}

In Fig.\ref{fig:3TXcoop} we simulate a $3$ TX cooperation system. The rate constraint on coordination link from TX~$2$ to TX~$1$ and TX~$3$ to TX~$1$ satisfies $R_{21}=R_{31}$. In this simulation, the parameters are denoted below:
\begin{align*}
&K\!=\!L\!=\!3, M\!=\!N\!=\!1, 2n\!=\!18,M_{2n}\!=\!\frac{1}{2\pi e},\mathbf{h}\!\sim\!\mathcal{CN}(\mathbf{0},\mathbf{I}_9)\\
&Q_1=diag(0.1,0.5,0.5,1,1,1,1,1,1)\\
&Q_2=diag(1,1,1,0.5,0.1,0.5,1,1,1)\\
&Q_3=diag(1,1,1,1,1,1,0.5,0.5,0.1)
\end{align*}
This simulation setting corresponds to the following case: each TX~$i$ has more accurate CSI for the user channel $\mathbf{H}_{ii}$, it also has some coarse CSI for the user channel $\mathbf{H}_{ji},\forall j\neq i$ and almost no CSI for the channel between the other TXs and the RXs. Fig.\ref{fig:3TXcoop} clearly shows the performance enhancement of the coordination shaping algorithm over the conventional unshaped coordination algorithm in a multiple TX cooperation scenario.

\begin{figure}[tp!]
\centering
\includegraphics[width=1\columnwidth]{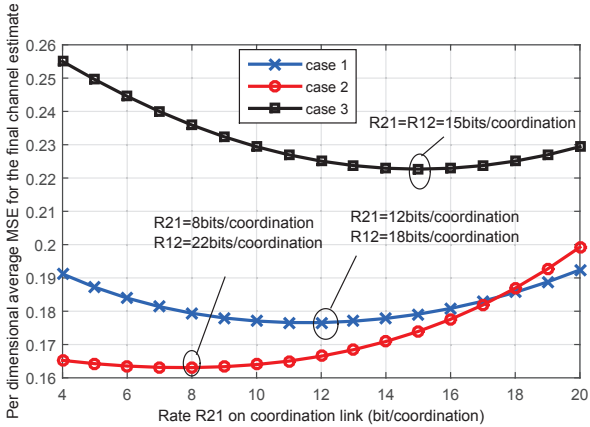}
\caption{Per dimensional average MSE for the final channel estimation at TX~$1$, TX~$2$ when sum bit for coordination link $R_{tot}=R_{21}+R_{12}=30$bits. Case $1$: $\mathbf{Q}_1=diag(0.1,0.1,0.9,0.9)$, $\mathbf{Q}_2=diag(0.5,0.5,0.5,0.5)$,
case $2$: $\mathbf{Q}_1=diag(0.4,0.2,0.3,0.1)$, $\mathbf{Q}_2=diag(0.7,0.8,0.6,0.9)$, case $3$: $\mathbf{Q}_1=diag(0.5,0.5,0.5,0.5)$, $\mathbf{Q}_2=diag(0.5,0.5,0.5,0.5)$.
}\label{fig:mseplot1}
\end{figure}

Fig.\ref{fig:mseplot1} considers the coordination link bit allocation problem for a $2$ TX cooperation system. The total amount of bits for coordination link is $R_{12}+R_{21}=30$bits/coordination. The figure reveals that the cooperation information exchange is not necessarily symmetric. In case $3$, the two TXs exchange information with equal rate $R_{12}=R_{21}=15$bits/coordination because the accuracy of CSI at both end is the same. However, in case $2$, the optimal coordination link bit allocation strategy is to let TX~$1$ share the cooperation information to TX~$2$ through a $R_{21}=8$bits/coordination backhaul and vice versa through a backhaul of $R_{12}=22$bits/coordination. It's intuitive because for every channel coefficient, TX~$1$ has a more accurate initial CSI than TX~$2$. It reveals that if one TX has a better CSI, it is more encouraged to share his information with its cooperating end through the coordination link.

\section{Conclusion}
\label{sec:conc}
We study decentralized cooperative channel estimation for use in transmitter coordinated systems under strict backhaul information exchange rate
limitations. We relate this problem to the information theoretic setup of distributed source coding with side information.
We derive a practical algorithm which near-optimally combines  finite-rate backhaul exchange together with local pre-existing CSI  at the transmitters. The scheme is robust with respect to arbitrary feedback noise statistics and allows to jointly optimize the quantization step used over the rate-limited backhaul links together with the weights using to combine local CSI with exchanged CSI.

\appendices
\section{Proof of Proposition \ref{coro4}}\label{sec:appenA}
Adopt the notation in Proposition \ref{coro4}, according to \eqref{eq:distortion}, \eqref{zki} and \eqref{finalest}, the per dimensional MSE can be expressed as:
\begin{align*}
D^{(i)}&=\frac{1}{n}\mathbb{E}\{\parallel\mathbf{h}-\tilde{\mathbf{h}}^{(i)}\parallel^{2}\}\\
&=\!\frac{1}{n}\!\Tr\!\left(\!(\!\underset{k\in\mathcal{A}_i}{\sum}\!\mathbf{W}_{ki}\!\mathbf{A}_{ki}\!+\!\mathbf{W}_{ii}\!-\!\mathbf{I}\!)\mathbf{Q}_{\mathbf{h}}(\!\underset{k\in\mathcal{A}_i}{\sum}\!\mathbf{W}_{ki}\!\mathbf{A}_{ki}\!+\!\mathbf{W}_{ii}\!-\!\mathbf{I}\!)^H\!\right)\\
&+\!\frac{1}{n}\!\Tr\!\left(\!\underset{k\in\mathcal{A}_i}{\sum}\mathbf{W}_{ki}\!\left(\!\mathbf{A}_{ki}\!\mathbf{Q}_{k}\!\mathbf{A}_{ki}^{H}\!+\!\mathbf{A}_{ki}\!\mathbf{Q}_{\mathcal{Q}_{ki}}\!\right)\!\mathbf{W}_{ki}^{H}\!+\!\mathbf{W}_{ii}\!\mathbf{Q}_{i}\!\mathbf{W}_{ii}^{H}\!\right).
\end{align*}
Take the partial derivatives and set them to zero:
\begin{align*}
&\frac{\partial D^{(i)}}{\partial\mathbf{W}_{ii}^{*}}=0\\
&\frac{\partial D^{(i)}}{\partial\mathbf{W}_{ki}^{*}}=0,\qquad \forall k\in\mathcal{A}_i,
\end{align*}
which leads to:
\begin{align*}
&\mathbf{W}_{ii}\left(\mathbf{Q}_{\mathbf{h}}+\mathbf{Q}_{i}\right)=\left(\mathbf{I}-\underset{k\in\mathcal{A}_i}\sum\mathbf{W}_{ki}\mathbf{A}_{ki}\right)\mathbf{Q}_{\mathbf{h}}\\
&\mathbf{W}_{ki}\left(\mathbf{A}_{ki}\mathbf{Q}_{\mathbf{h}}\mathbf{A}_{ki}^H+\mathbf{A}_{ki}\mathbf{Q}_k\mathbf{A}_{ki}^{H}+\mathbf{A}_{ki}\mathbf{Q}_{\mathcal{Q}_{ki}}\right)\\
&=\left(\mathbf{I}-\underset{\substack{t\in\mathcal{A}_i\\t\neq k}}{\sum}\mathbf{W}_{ti}\mathbf{A}_{ti}-\mathbf{W}_{ii}\right)\mathbf{Q}_{\mathbf{h}}\mathbf{A}_{ki}^H.
\end{align*}
Solve the above equation system, the optimal weight combing matrices $\{\mathbf{W}_{ki}^{opt},\mathbf{W}_{ii}^{opt},k\in\mathcal{A}_i\}$ can be derived as:
\begin{align*}
&\left[\begin{array}{cccc}
\mathbf{W}_{ii}^{opt} & \mathbf{W}_{l_1i}^{opt} & \ldots & \mathbf{W}_{l_{C_i}i}^{opt}
\end{array}\right]
=\mathbf{Q}_{\mathbf{h}}\bm\Upsilon_i\bm\Omega_i^{-1}.\\
\end{align*}
Let
\begin{align*}
\mathbf{W}&=\left[\begin{array}{cccc}
\mathbf{W}_{ii}^{opt} & \mathbf{W}_{l_1i}^{opt} & \ldots & \mathbf{W}_{l_{C_i}i}^{opt}
\end{array}\right]\\
\bm\Theta_i&=\!\left[\!\begin{array}{cccc}
\mathbf{Q}_i & 0 & \ldots & 0\\
0 & \mathbf{P}_{l_1i}\!-\!\mathbf{A}_{l_1i}\!\mathbf{Q}_{\mathbf{h}}\!\mathbf{A}_{l_1i}^H & \ldots & 0\\
\vdots & \vdots & \ddots & \vdots\\
0 & 0 & \ldots & \mathbf{P}_{l_{C_i}i}\!-\!\mathbf{A}_{l_{C_i}i}\!\mathbf{Q}_{\mathbf{h}}\!\mathbf{A}_{l_{C_i}i}^H
\end{array}\!\right],
\end{align*}
then
\begin{align*}
\bm\Omega_i=\bm\Theta_i+\bm\Upsilon_i^H\mathbf{Q}_{\mathbf{h}}\bm\Upsilon_i.
\end{align*}
Since
\begin{align*}
\mathbf{W}=\mathbf{Q}_{\mathbf{h}}\bm\Upsilon_i\bm\Omega_i^{-1},
\end{align*}
the optimum per dimensional MSE satisfies
\begin{align*}
D^{(i)opt}&=\frac{1}{n}\Tr\left((\mathbf{W}\bm\Upsilon_i^H-\mathbf{I})\mathbf{Q}_{\mathbf{h}}(\mathbf{W}\bm\Upsilon_i^H-\mathbf{I})^H+\mathbf{W}\bm\Theta_i\mathbf{W}^H\right)\\
&=\frac{1}{n}\Tr\left(\mathbf{Q}_{\mathbf{h}}-\mathbf{Q}_{\mathbf{h}}\bm\Upsilon_i\bm\Omega_i^{-1}\bm\Upsilon_i^H\mathbf{Q}_{\mathbf{h}}\right)\\
&\stackrel{(a)}{=}\frac{1}{n}\Tr\left(\mathbf{Q}_{\mathbf{h}}^{-1}+\bm\Upsilon_i\bm\Theta_i^{-1}\bm\Upsilon_i^H\right)^{-1}\\
&=\frac{1}{n}\!\Tr\!\left(\!\mathbf{Q}_{\mathbf{h}}^{-1}\!+\!\!\!\underset{k\in\mathcal{A}_i}{\sum}\!\mathbf{A}_{ki}^H\!(\!\mathbf{P}_{ki}\!-\!\mathbf{A}_{ki}\!\mathbf{Q}_{\mathbf{h}}\!\mathbf{A}_{ki}^H\!)^{-1}\mathbf{A}_{ki}\!+\!\mathbf{Q}_i^{-1}\!\right)^{\!-1}\\
&=\frac{1}{n}\Tr\left(\mathbf{Q}_{\mathbf{h}}^{-1}+\mathbf{Q}_i^{-1}+\bm\Lambda_i\right)^{-1},
\end{align*}
where $(a)$ follows from the Woodbury identity
\begin{align*}
(\mathbf{A}\!+\!\mathbf{C}\mathbf{B}\mathbf{C}^H)^{-1}\!=\!\mathbf{A}^{-1}\!-\!\mathbf{A}^{-1}\mathbf{C}(\mathbf{B}^{-1}\!+\!\mathbf{C}^H\mathbf{A}^{-1}\mathbf{C})^{-1}\mathbf{C}^H\mathbf{A}^{-1}.
\end{align*}
This concludes the proof.

\section{Proof of Proposition \ref{coro1}}\label{sec:appenB}
In order to prove this proposition, we need the following lemma:
\newtheorem{lemma3}{Lemma}
\begin{lemma3}\label{l3}
for a random vector $\mathbf{x}$, if an optimal $S$ levels Euclidean distance distortion based quantizer applied on a random vector $\mathbf{y}=\mathbf{B}^{\frac{1}{2}}\mathbf{x}$ has the codebook $\{\mathbf{y}_1,\dots,\mathbf{y}_S\}$ and associated partition $\{\mathcal{P}_1,\dots,\mathcal{P}_S\}$, then the optimal $S$ level weighted square error distortion based quantizer applied on the random vector $\mathbf{x}$ with shaping matrix $\mathbf{B}$ will have the codebook $\{\mathbf{B}^{-\frac{1}{2}}\mathbf{y}_1,\dots,\mathbf{B}^{-\frac{1}{2}}\mathbf{y}_S\}$ and associated partition $\{\mathbf{B}^{-\frac{1}{2}}[\mathcal{P}_1],\dots,\mathbf{B}^{-\frac{1}{2}}[\mathcal{P}_S]\}$,where the $\mathbf{B}^{-\frac{1}{2}}[\mathcal{P}_i]$ is defined as $\mathbf{B}^{-\frac{1}{2}}[\mathcal{P}_i]=\{\mathbf{x}:\exists\mathbf{y}\in\mathcal{P}_i\textrm{ s.t. }\mathbf{x}=\mathbf{B}^{-\frac{1}{2}}\mathbf{y}\}$.
\end{lemma3}
\begin{proof}
Consider the distortion associated with the optimal codebook and partition:
\begin{align*}
D_{\textbf{y}}&=\underset{j=1}{\overset{S}\sum}f_{\textbf{y}}(\textbf{y}_j)\underset{\textbf{y}\in \mathcal{P}_j}\int(\textbf{y}-\textbf{y}_j)^H(\textbf{y}-\textbf{y}_j)d\textbf{y}.\\
D_{\textbf{x}}&=\underset{j=1}{\overset{S}\sum}f_{\textbf{x}}(\textbf{x}_j)\underset{\textbf{x}\in \mathcal{M}_j}\int(\textbf{x}-\textbf{x}_j)^HB(\textbf{x}-\textbf{x}_j)d\textbf{x}\\
&=\det(\textbf{B}^{-1})\underset{j=1}{\overset{S}\sum}f_{\textbf{x}}(\textbf{x}_j)\underset{\textbf{u}\in \textbf{B}^{\frac{1}{2}}[\mathcal{M}_j]}\int(\textbf{u}-\textbf{B}^{\frac{1}{2}}\textbf{x}_j)^H(\textbf{u}-\textbf{B}^{\frac{1}{2}}\textbf{x}_j)d\textbf{u},
\end{align*}
where $\{\textbf{x}_1,\dots,\textbf{x}_S\}$ and $\{\mathcal{M}_1,\dots,\mathcal{M}_S\}$ denote the optimal codebook and partition for the quantizing $\textbf{x}$.
Let $\textbf{B}^{\frac{1}{2}}\textbf{x}_j=\textbf{y}_j$ and $\textbf{B}^{\frac{1}{2}}[\mathcal{M}_j]=\mathcal{P}_j$, since
\begin{align*}
f_{\textbf{y}}(\textbf{y}_j)=f_{\textbf{x}}(\textbf{x}_j)\det(\textbf{B}^{-1}),
\end{align*}
by change of variable in the integral, we can get $D_{\textbf{y}}=D_{\textbf{x}}$.
\end{proof}
Thus, according to Lemma \ref{l3},
\begin{eqnarray*}
\mathbf{Q}_{\mathbf{x}}
&=&\mathbb{E}\{(\mathbf{x}-\mathcal{Q}(\mathbf{x}))(\mathbf{x}-\mathcal{Q}(\mathbf{x}))^H\}\\
&=&\mathbf{B}^{-\frac{1}{2}}\mathbb{E}\{(\mathbf{y}-\mathcal{Q}(\mathbf{y}))(\mathbf{y}-\mathcal{Q}(\mathbf{y}))^H\}\mathbf{B}^{-\frac{H}{2}}\\
&=&\mathbf{B}^{-\frac{1}{2}}\mathbf{Q}_{\mathbf{y}}\mathbf{B}^{-\frac{H}{2}},
\end{eqnarray*}
where $\mathbf{y}=\mathbf{B}^{\frac{1}{2}}\mathbf{x}\sim\mathcal{CN}(\mathbf{0},\mathbf{B}^{\frac{1}{2}}\bm\Gamma\mathbf{B}^{\frac{H}{2}})$.
Let $\mathbf{t}=[\Re(\mathbf{y})^T \Im(\mathbf{y})^T ]^T$, then $\mathbf{t}\sim\mathcal{N}(\mathbf{0},\bm\Phi)$ and
\[
\bm\Phi=\frac{1}{2}\left[
\begin{array}{lcl}
\Re(\mathbf{B}^{\frac{1}{2}}\bm\Gamma\mathbf{B}^{\frac{H}{2}}) & \Im(\mathbf{B}^{\frac{1}{2}}\bm\Gamma\mathbf{B}^{\frac{H}{2}})\\
\Im(\mathbf{B}^{\frac{1}{2}}\bm\Gamma\mathbf{B}^{\frac{H}{2}}) & \Re(\mathbf{B}^{\frac{1}{2}}\bm\Gamma\mathbf{B}^{\frac{H}{2}})
\end{array}\right].
\]
Furthermore, it is proven in \cite{508838} that
\[
\mathbf{Q}_{\mathbf{t}}=\mathbb{E}\{(\mathbf{t}-\mathcal{Q}(\mathbf{t}))(\mathbf{t}-\mathcal{Q}(\mathbf{t}))^T\}=D_{\mathbf{t}}\mathbf{I}_n,
\] where
the average distortion $D_{\mathbf{t}}=\frac{1}{n}\Tr(\mathbf{Q}_{\mathbf{t}})$ is obtained from \cite{1056490} for large $S$ as
\begin{eqnarray*}
D_{\mathbf{t}}
&=&S^{-\frac{2}{n}}M_n\left(\int f_{\mathbf{t}}(\mathbf{t})^{\frac{n}{n+2}}d\mathbf{t}\right)^{\frac{n+2}{n}}\\
&=&S^{-\frac{2}{n}}M_n2\pi\left(\frac{n+2}{n}\right)^{\frac{n}{2}+1}\det(\bm\Phi)^{\frac{1}{n}},
\end{eqnarray*}
where $f_{\mathbf{t}}(.)$ is the probability density function (p.d.f) of $\mathbf{t}$.
Finally, from the expression of $\mathbf{Q}_{\mathbf{t}}$ and by a real-to-complex conversion, we get
\[
\mathbf{Q}_{\mathbf{y}}=2S^{-\frac{1}{n}}M_{2n}2\pi(\frac{n+1}{n})^{n+1}\det(\bm\Phi)^{\frac{1}{2n}}\mathbf{I}_{n},
\]
which leads to
\begin{eqnarray*}
\mathbf{Q}_{\mathbf{x}}&=&2S^{-\frac{1}{n}}M_{2n}2\pi(\frac{n+1}{n})^{n+1}\det(\bm\Phi)^{\frac{1}{2n}}\mathbf{B}^{-1}\\
&=&S^{-\frac{1}{n}}M_{2n}2\pi(\frac{n+1}{n})^{n+1}\det(\mathbf{B})^{\frac{1}{n}}\det(\bm\Gamma)^{\frac{1}{n}}\mathbf{B}^{-1}\\
&=&\mathbf{Q}_{0}^{(S)}(\bm\Gamma)\det(\mathbf{B})^{\frac{1}{n}}\mathbf{B}^{-1},
\end{eqnarray*}
where
\begin{equation*}
\mathbf{Q}_{0}^{(S)}(\bm\Gamma)=S^{-\frac{1}{n}}M_{2n}2\pi\left(\frac{n+1}{n}\right)^{n+1}\det(\bm\Gamma)^{\frac{1}{n}}\mathbf{I}_n.
\end{equation*}

\section{Proof of Proposition \ref{coro5}}\label{sec:appenC}
We first give some necessary lemmas:
\newtheorem{lemma2}[lemma3]{Lemma}
\begin{lemma2}\label{l2}
\cite[Lemma 2.5]{brinkhuis2005matrix} The matrix function $h:\mathbf{X}\mapsto-\mathbf{X}^{-1}$ is both a strictly matrix concave function and a matrix monotone increasing function.
\end{lemma2}

\newtheorem{lemma1}[lemma3]{Lemma}
\begin{lemma1}\label{l1}
If $\mathbf{X}$ is a positive semi-definite hermitian matrix, then matrix function $h:\mathbf{X}\mapsto\Tr(\mathbf{X}^{-1})$ is convex and matrix monotone decreasing.
\end{lemma1}
\begin{proof}
Consider $g(t)=h(\mathbf{Z}+t\mathbf{V})$, where $\mathbf{Z}$ is positive semi-definite hermitian matrix, $\mathbf{V}$ is hermitian matrix, since
\begin{align*}
\frac{d^2}{dt^2}g(t)|_{t=0}=\Tr\left(2(\mathbf{Z}^{-1}\mathbf{V})\mathbf{Z}^{-1}(\mathbf{Z}^{-1}\mathbf{V})^{-1}\right)\geq0,
\end{align*}
therefore $f(\mathbf{X})$ is convex.\\
For $\mathbf{X}\succeq\mathbf{Y}\succeq\mathbf{0}$, according to Lemma \ref{l2},
\begin{align*}
&\mathbf{Y}^{-1}\succeq\mathbf{X}^{-1}\succeq\mathbf{0}\\
&h(\mathbf{X})-h(\mathbf{Y})=\Tr(\mathbf{X}^{-1}-\mathbf{Y}^{-1})\leq0,
\end{align*}
therefore, the function is matrix monotone decreasing.
\end{proof}

Adopt the notations in Proposition \ref{coro4}, we will prove the convexity of problem (\ref{optimizeB}). Since
\begin{align*}
\det(\mathbf{B}_{ki})=1,
\end{align*}
according to \eqref{Qquantki},
\begin{align*}
\mathbf{Q}_{\mathcal{Q}_{ki}}=\mathbf{Q}_{0}^{(S)}(\bm\Gamma)\mathbf{B}_{ki}^{-1}.
\end{align*}
Let
\begin{align*}
f=\mathbf{Q}_{\mathbf{h}}^{-1}+\mathbf{Q}_i^{-1}+\bm\Lambda_i,
\end{align*}
with $\bm\Lambda_i$ defined in Corollary \ref{coro4}. Let's consider the convexity of composite function: if a function $h$ is concave and matrix monotone increasing, a function $g$ is a concave function, then the composite function $h\circ g$ is concave. According to Lemma \ref{l2},
\begin{align*}
h=-\mathbf{X}^{-1},
\end{align*}
is matrix concave and matrix monotone increasing for $\mathbf{X}$. According to Lemma \ref{l2},
\begin{align*}
g=\mathbf{A}-\mathbf{B}_{ki}^{-1},
\end{align*}
with constant matrix $\mathbf{A}$ is concave for $\mathbf{B}_{ki}$. Therefore the functions
\begin{align*}
\bm\Lambda_i&=h\circ g\\
f&=\mathbf{Q}_{\mathbf{h}}^{-1}+\mathbf{Q}_i^{-1}+\bm\Lambda_i
\end{align*}
are concave for $\mathbf{B}_{ki}$.

Consider the convexity of composite function: if a function $h$ is convex and matrix monotone decreasing, a function $g$ is concave function, then the composite function $h\circ g$ is convex. According to Lemma \ref{l1},
\begin{align*}
h=\Tr\left(\mathbf{X}^{-1}\right)
\end{align*}
is convex and matrix monotone decreasing for $\mathbf{X}$, since
\begin{align*}
g&=\mathbf{Q}_{\mathbf{h}}^{-1}+\mathbf{Q}_i^{-1}+\bm\Lambda_i
\end{align*}
is concave, therefore we can conclude that
\begin{align*}
D^{(i)opt}=h\circ g
\end{align*}
is convex for $\mathbf{B}_{ki}$.

When the coordination link rate $R_{ki}$ is sufficiently large, use twice matrix inverse approximation:
\begin{align*}
D^{(i)opt}&=\frac{1}{n}\Tr\left(\mathbf{Q}_{\mathbf{h}}^{-1}+\mathbf{Q}_i^{-1}+\bm\Lambda_i\right)^{-1}\\
%&=\!\frac{1}{n}\!\Tr\!\!\left[\!\mathbf{Q}_{\mathbf{h}}^{-1}\!+\!\mathbf{Q}_i^{-1}\!+\!\!\underset{k\in\mathcal{A}_i}{\sum}\!\!\left(\!(\!\mathbf{Q}_{\mathbf{h}}\!+\!\mathbf{Q}_k\!)\!(\!\mathbf{Q}_{\mathbf{h}}\!+\!\mathbf{Q}_k\!-\!\mathbf{Q}_{\mathcal{Q}_{ki}}\!)^{\!-1}\!(\!\mathbf{Q}_{\mathbf{h}}\!+\!\mathbf{Q}_k\!)\!-\!\mathbf{Q}_{\mathbf{h}}\!\right)^{\!-1}\!\right]^{\!-1}\\
&\stackrel{(a)}{\simeq}\frac{1}{n}\Tr\left(\underset{k\in\mathcal{A}_i}{\sum}\left(\mathbf{Q}_k+\mathbf{Q}_{\mathcal{Q}_{ki}}\right)^{-1}+\mathbf{Q}_i^{-1}+\mathbf{Q}_{\mathbf{h}}^{-1}\right)^{-1}\\
&\stackrel{(b)}{\simeq}\frac{1}{n}\!\Tr\!\left(\underset{k\in\mathcal{A}_i}{\sum}\!(\mathbf{Q}_k^{-1}\!-\!\mathbf{Q}_k^{-1}\mathbf{Q}_{\mathcal{Q}_{ki}}\mathbf{Q}_k^{-1})\!+\!\mathbf{Q}_i^{-1}\!+\!\mathbf{Q}_{\mathbf{h}}^{-1}\right)^{-1},
\end{align*}
where $(a)$ and $(b)$ follows from the matrix inverse first order approximation: if $\mathbf{X}^n\rightarrow0$ when $n\rightarrow\infty$, then
\begin{align*}
(\mathbf{A}+\mathbf{X})^{-1}\simeq\mathbf{A}^{-1}-\mathbf{A}^{-1}\mathbf{X}\mathbf{A}^{-1}.
\end{align*}
Similar to the previous analysis, according to convexity for composite matrix function, applying Lemma \ref{l1} and Lemma \ref{l2}, we can conclude the above approximation for $D^{(i)opt}$ is convex.

If we relax the constraint $\det(\mathbf{B}_{ki})=1$ to $\det(\mathbf{B}_{ki})\geq1$, the feasible set for $\mathbf{B}_{ki}$ is therefore a convex set and the optimization for this relaxed problem is minimizing a convex function over a convex set, which is a convex optimization problem and the optimal $\mathbf{B}_{ki}$ should attain at the boundary $\det(\mathbf{B}_{ki})=1$.This concludes the proof.
\section*{Acknowledgment}

D. Gesbert is partially supported by the ERC under the European Union's Horizon 2020 research and innovation program (Agreement no. 670896).
The authors would like to thank Dr. Paul de Kerret, the editor and anonymous reviewers for their valuable suggestions to improve the work.

\ifCLASSOPTIONcaptionsoff
  \newpage
\fi



%\bibliographystyle{IEEEtran}
\bibliographystyle{IEEEtran}
\bibliography{Bibliography}

\end{document}